\documentclass[12pt,amsfonts]{article}    

\usepackage{graphicx}

\def\vp{\varphi}

\def\half{\textstyle{\frac{1}{2}}}

\def\half{\textstyle{\frac{1}{2}}}

\def\H{{\cal H}}

\def\H{{\cal H}}
\def\G{{\cal{G}}}

\def\da{{\bf{a}}}

\def\l{\lambda}

\def\ra{\rightarrow}
\def\tint{{\textstyle\int}}

\def\d{\partial}

\def\b{\begin{eqnarray*}}  
\def\e{\end{eqnarray*}}    
\def\bn{\begin{eqnarray}}  
\def\en{\end{eqnarray}}   

\def\<{\langle}
\def\lp{\overline{p}}
\def\lH{\overline{H}}
\def\bS{{\bf{S}}}
\def\lq{\overline{q}}
\def\>{\rangle}

\def\no{\nonumber}

\def\{{\lbrace}
\def\}{\rbrace}
\def\W{{\cal{W}}}
\def\om{\omega}

\begin{document}  

\title{A Unified Combination of \\ Classical and Quantum Systems}        
\author{John R. Klauder\footnote{klauder@phys.ufl.edu} \\
Department of Physics and Department of Mathematics \\
University of Florida,   
Gainesville, FL 32611-8440}
\date{ }
\bibliographystyle{unsrt}
\maketitle

\begin{abstract}Any particular classical system and its quantum version are normally
viewed as separate formulations that are strictly distinct. Our goal is to overcome the two separate languages and create a smooth and common procedure that provides a clear and continuous passage between the conventional distinction of either a strictly classical or a strictly quantized state. While path integration, among other procedures, provides an alternative route to connect classical and quantum expressions, it normally involves complicated, model-dependent, integrations. Our alternative procedures involve only model-independent procedures, and use more natural and
straightforward integrations that are universal in kind. To introduce the basic procedures our presentation begins with familiar methods that are limited to basic, conventional, canonical quantum mechanical examples. In the final sections we illustrate how alternative quantization procedures, e.g., spin and affine quantizations, can also have smooth paths between classical and quantum stories, and with a few brief remarks, can also lead to similar stories for non-renormalizable covariant scalar fields as well as quantum gravity. 
\end{abstract}
\section{Introduction}
\subsection{Classical formulations}
We start by describing phase space as a set of momenta ($p$) and coordinates ($q$), which have a
measure ($dp\,dq$). Dynamical equations are time ($t$) dependent ($p(t)$) and ($q(t)$), which are dictated by 
stationary variations of a classical ($c$) action functional 
  \bn A_c=\tint_0^T \{ p(t)\,\dot{q}(t)- H(p(t),q(t))\,\}\;dt\;, \en
  that leads, with $\dot{q}(t)\equiv d\,q(t)/dt$, to 
  \bn &&\delta A_c=\tint_0^T \{ [\dot{q}(t)-\d H(p(t),q(t)/\d p(t)]\,\delta  p(t)\no \\
  &&\hskip5em -[\dot{p}(t)+\d H(p(t),q(t))/\d q(t)]\,\delta q(t)\}\;dt \no \\
  &&\hskip7em+q(t)\,\delta p(t)|_0^T-p(t)\,\delta q(t)|_0^T=0 \;, \en
  with arbitrary  $\delta p(t)$ and $\delta q(t)$ values, except that $\delta p(T)=\delta p(0)=0$
  and $\delta q(T)=\delta q(0)=0$. All this leads to the traditional classical equations of motion given by
    \bn \dot{q}(t)&&\hskip-1,3em =\d \,H(p(t), q(t))/\d p(t)\;, \no \\
         \dot{p}(t)&&\hskip-1.3em=-\d\,H(p(t),q(t))/\d q(t)\;. \en
         
  A common example is $H(p,q)=(p^2/m+m\,\om^2 q^2)/2$, which is an harmonic oscillator, and its dynamical equations 
  are given by $\dot{q}(t) = p(t)/ m$ and $\dot{p}(t)=-m\,\om^2 q(t)$. It follows that $\ddot{q}(t)=-\om^2 q(t)$
  with solutions $q(t)= q_0\cos(\om\,t)+(p_0/m\,\om)\sin(\om\,t)$, where $q_0=q(0)$ 
  and $p_0= p(0)$.
         
 \subsection{Quantum formulations}
 The basic quantum operators are linked with suitably chosen classical variables, i.e., $p\ra P$ and $q\ra Q$, which satisfy the rule $QP-PQ\equiv [Q,P]=i\hbar 1\!\!1$. {\bf Note:} The rules used to choose the favored classical variables, $p$ and $q$, that are 
 promoted to operators, $P$ and $Q$, are addressed below.
 
 Various operators, such as $\W(P,Q)$, act upon suitable Hilbert space vectors, such as $|\Phi\>$ leading to new Hilbert space vectors, $\W(P,Q)|\Phi\>$. 
 It is conventionally observed that the operator $Q$ is diagonalized when acting on selected
 eigenvectors. This leads to the relations that $Q|x\>=x|x\>$ while then $P|x\>=-i\hbar(\d/\d x) \,|x\>$. A general Hilbert space vector $|\Phi\>$ then becomes $\Phi(x)=\<x|\Phi\>$, and the normalization of a general vector is $\<\Phi|\Phi\>=\tint |\Phi(x)|^2\;dx<\infty$. While such relations formally are correct, it should be recognized that $|x\>$ is strictly not a proper Hilbert space vector since
 $\<x|x\>=\infty$ as follows from the fact that $\<x'|x\>=\delta(x'-x)$.

 The algebra of quantum operators, such as the self-adjoint Hamiltonian operator $\H(P,Q)$, act upon vectors from a suitable Hilbert space. Such vectors as
 $|\Phi\>$ are finite in normalization, as denoted by $\<\Phi|\Phi\><\infty$. Indeed, normalized vectors, which means that
 $\<\Phi|\Phi\>=1$, play an important role. Quantum dynamics is expressed by time dependence of the vectors, e.g., $|\Phi(t)\>$. The quantum ($q$) action functional involves normalized vectors, and is given by
    \bn A_q=\tint_0^T  \<\Phi(t)|[i\hbar\,(\d/\d t)- \H(P,Q)]|\Phi(t)\>\;dt\;, \label{tt}\en
 and stationary variations of the quantum action functional lead to
    \bn &&\delta A_q=\tint_0^T \{\delta\<\Phi(t)|[i\hbar\,(\d/\d t)- \H(P,Q)]|\Phi(t)\>\; \no\\
         &&\hskip5em+\<\Phi(t)|[i\hbar\,(\d/\d t)- \H(P,Q)]\,\delta |\Phi(t)\>\,\}\;dt=0\;.\en
         This leads to two equations, one of which is
         \bn i\hbar (\d \,|\Phi(t)\>/\d t)=\H(P,Q)\,|\Phi(t)\>\;, \en
         which is a version of `Schr\"odinger's equation' \cite{1}; the second equation is the adjoint of the first equation.

 The basic operators, $P$ and $Q$, can generate various additional operators, such as an harmonic oscillator given by  $\H(P,Q)=(P^2/m+m\,\om^2 Q^2)/2$. For the
 Hilbert space `vectors'  $|x\>$, defined by $Q|x\>=x|x\>$, the ground-state eigenvector of the harmonic oscillator is $\<x|\om\>={\cal{N}}\exp[-m\,\om \,x^2/2\hbar]$, i.e., namely a familiar Gaussian function.

\section{Coherent States}
\subsection{Choosing the correct quantum operators}
Since classical mechanics has a long history, it is natural that the classical variables are
chosen before choosing what quantum operators to employ. That behavior has its natural difficulties because while there are many acceptable classical variables $p$ and $q$, or $\lp$ and $\lq$, etc., provided the Hamiltonian function also is changed so that $\lp=\lp(p,q)$ and $\lq=\lq(p,q)$, along with the Poisson brackets $\{p,q\}=\{\lp,\lq\}=1$, which leads to $\lH(\lp,\lq)=H(p,q)$ for all variables.

On the other hand, a promotion of classical variables to quantum operators, such as 
$p\ra P$ and $q\ra Q$, {\it but generally}, $\overline{H}(\overline{P},\overline{Q})\neq H(P,Q)$,
and when the Hamiltonian operators are different, they definitely can lead to different physics. It is {\it absolutely necessary} to find a procedure whereby {\it favored} classical variables are the ones that are promoted to the {\it correct quantum operators}. Accepting an arbitrary choice of
canonical variables to promote to quantum operators would most likely lead to a {\it false quantization procedure}. How a correct choice of canonical variables to promote to quantum operators can be made is the subject of this subsection.

Dirac \cite{2} has offered the clue to the task ahead. He claimed that the favored classical
variables, $p$ and $q$, are those that obey the relation $H(p,q)\ra H(P,Q)$, i.e., the classical functional form equals the quantum functional form, and moreover, the classical variables must also be Cartesian coordinates, a rule that necessarily implies that $-\infty<p,q<\infty$. 
At the first glance this seems impossible because phase space $-$ the home of the variables $p$ and 
$q$ $-$ has no metric to determine Cartesian coordinates.
Dirac did not propose how to find such coordinates, but the author has recently found a suitable procedure to do just that. That story is published in \cite{3}, and is recast here below.

To seek Cartesian coordinates we first choose a family of canonical coherent states. We start by choosing $P$ and $Q$, such that $[Q,P]=i\hbar1\!\!1$, and a pair of classical variables $p$ and $q$. For our coherent states we choose Hilbert space vectors such as
    \bn |p,q\> \equiv e^{-iqP/\hbar}\,e^{ipQ/\hbar}|\omega\>\;, \en
where it is customary that $(Q+iP/\omega)\,|\omega\>=0$; this choice will also play an important role later. It follows that $\<\om|(Q+iP/\om)|\om\>=0$, which implies that $\<\om|P|\om\>=\<\om|Q|\om\>=0$.
    
    The coherent states span the Hilbert space as noted by the resolution of the identity, $1\!\!1$, given by \bn \int |p,q\>\<p,q| \: dp\,dq/2\pi\hbar =1\!\!1 \;. \label{rrr}\en 
     This expression will also have a role to play in our following analysis.
    
    At this stage, it is useful to assume $Q$ (and $q$) are dimensionless so that $P$ (and $p$), and also $\om$, have the dimensions of $\hbar$. Next, a phase factor is added to
    the coherent states, which will be useful in what follows. Specifically, we add the phase factor to the coherent states as $|p,q:F\>\equiv e^{iF(p,q,b)}\,|p,q\>$. Note that no operators appear in the real function $F(p,q,b)$ whose variables, $p$ and $q$, and any other variable(s) $b$, can lead to an arbitrary phase modification of the coherent states.

    We next maintain that a special semi-classical connection exists between selected
   classical and quantum expressions, and which, for clarity, our expressions are assumed to be polynomial Hamiltonians, given by
     \bn &&H(p,q)=\<p,q| \H(P,Q)|p,q\>\no \\
          &&\hskip3.5em =\<\om| \H(P+p,Q+q)|\om\> \no \\
          &&\hskip3.5em=\H(p,q)+ {\cal{O}}(\hbar;p,q) \;. \label{ee} \en
          Observe first that each of the three lines of expression (\ref{ee}) are {\it identical} if one exchanges the coherent states vectors $|p,q\>$ for any of the phase modified coherent state vectors,  $|p,q:F\>$. In the limit that $\hbar\ra 0$ $-$ which leads one to the true classical 
          level $-$ observe that the quantum function $\H$ equals the classical function $H$, just as Dirac required.
          
    To complete the story, we introduce a Fubini-Study (F-S) metric \cite{4}, which 
    features a differential of the coherent state vectors such that they involve rays instead of vectors, a property which 
    means that the coherent state vectors belong in ray sets which are completely insensitive to any      phase factor $F(p,q,b)$ and, effectively, are such that $F( p,q,b)\equiv 0$. When the vectors
  $|p,q;F\> =e^{iF(p,q,b)}|p,q\>$ become completely independent of $F(p,q,b)$, it implies that $F(p,q,b)$ carries no physics. In particular, we choose an F-S expression as a formula designed to be
  independent of $F(p,q,b)$ and given by
      \bn  d\sigma(p,q)^2\equiv 2\hbar
       [\,|\!|\,d|p,q\>\,|\!|^2 - |\<p,q|\,d|p,q\>|^2\,]=\om^{-1}\,dp^2+ \om\,dq^2 \;, \en
 which leads to the fact that $p$ and $q$ are Cartesian variables after all, as Dirac had sought.  Thus, for this example, we have confirmed that $p$ and $q$ are the favored classical variables to promote to the physically correct quantum operators, $P$ and $Q$.\footnote{The paper \cite{aa}
 is devoted to how to choose favored classical variables for systems for which Cartesian coordinates are appropriate as well as other systems where Cartesian coordinates are  not appropriate.}
 
\subsection{The quantum action functional \\ restricted to coherent states} 
In (\ref{tt}) we have outlined the quantum action functional. Its stationary variations lead to the
Schr\"odinger equation in its abstract form. This equation is a fundamental relation in the quantum
story.  However, as classical observers, we are not able to vary all quantum vectors, but rather to a subset of vectors such as our family of coherent states. That proposal leads us to a semi-classical
relation, helped by $\<\omega|P|\omega\>=\<\omega|Q|\omega\>=0$, and given by
  \bn &&\hskip-2.4em A_{sc}=\int_0^T\{ \<p(t),q(t)|[\,i\hbar(\d/\d t)-\H(P,Q)\,]|p(t),q(t)\>\,\}\;dt \no \\ &&\hskip-.6em =\int_0^T\{\<\om|\dot{q}(t)(P+p(t))-\dot{p}(t) Q-\H(P+p(t),Q+q(t))|\om\>\,\}\;dt \no \\
    &&\hskip-.6em=\int_0^T \{ \, p(t)\,\dot{q}(t)-H(p(t),q(t))\,\}\;dt \;,  \label{jk} \en
  an equation that appears just like a classical action functional, but is only semi-classical in reality because $\hbar$ still has its normal positive value. 
  Even the dynamical equations that stationary variations lead to, specifically
   \bn &&\dot{q}(t)= \d H(p(t),q(t))/\d p(t) \no \\
        &&\dot{p}(t)=-\d H(p(t),q(t))/\d q(t) \;, \en
have to deal with $\hbar>0$. Stated briefly, the quantum story fits well into the semi-classical story.  The limit $\hbar\ra 0$    leads to the usual classical story in which $\hbar=0$. But $\hbar=0$ is {\it not} good physics because Mother Nature decided long ago that $\hbar>0$ so that atoms don't collapse and all the consequences that would imply. For further  discussion of this 
general topic, see \cite{5}.

As already noted, the expression $H(p,q)$ admits the relation $H(p,q)={\H}(p,q)+{\cal{O}}(\hbar;p,q)$, for which the    second term is generally 
ignored because  it is usually incredibly tiny; such a term is then dropped, and this action leads to the usual classical story, which then effectively pretends that $\hbar=0$.

These equations show that the quantum action passes smoothly to the classical action.
 We next ask if it would be possible to have the classical action pass smoothly to the quantum action; our surprising answer shows that it can be done!

\section{The Union of Classical and \\ Quantum Systems}
Having traced the semi-classical action functional to (\ref{jk}) we are in position to break the 
integrand being integrated in the top line into two separate  partners, 
namely $A=\<p(t),q(t)|\,i\hbar(\d/\d t)$ and $B=-\H(P,Q)\,|p(t),q(t)\>$. With this division we can even adopt two different times and thus choose the two vectors to be distinct, which, for example, leads to $A'= \<p,q| \,i\hbar (\d/\d t)$ and $B'=-\H(P,Q)\,|p',q'\>$, and now involves two different vectors.
Based on (\ref{rrr}) we can make use of two coherent state resolutions of the identity, namely
 $A'' =\tint\<\Phi(t)|p,q\>\<p,q| \,i\hbar(\d/\d t) \;dp\,dq/2\pi\hbar  =\<\Phi(t)|\,i\hbar(\d/\d t)$ and $B''=-\tint \H(P,Q) |p',q'\>\<p',q'|\Phi(t)\>\;dp'\,dq'/2\pi\hbar=-\H(P,Q)\,|\Phi(t)\>$.
 Finally, we restore the two parts together, i.e., $A''+B''$, and integrate the time $t$, which leads to the expression
   \bn A_q=\int_0^T \<\Phi(t)| [i\hbar(\d/\d t)-\H(P,Q)]|\Phi(t)\>\;dt\;, \en
   which is exactly the quantum action functional created by fleshing out the semi-classical
   expression in (\ref{jk}).
   
   \subsection{A bridge leading smoothly between the \\classical realm and the quantum realm}
   The purpose of cutting the semi-classical expression in (\ref{jk}) into two pieces was  to 
   officially 
   justify introducing two different coherent states. However, the reader should be willing to accept that $\<p,q|[i\hbar(\d/\d t)-{\H}(P,Q)]|p',q'\>$,  a purely mathematical expression that is not part of any classical or quantum elements, is an expression that can act as a `bridge', first to lead to `classical-land' by means of
     \bn \int\!\int \<p(t),q(t)|p,q\>\<p,q|[i\hbar(\d/\d t)- \H(P,Q)]|p',q'\> \no \\
     \times\<p',q'|p(t),q(t)\>\;dp\,dq\;dp'dq'/(2\pi\hbar)^2 \;,\en
     and second to lead to `quantum-land' by means of
     \bn \int\!\int \<\Psi(t)|p,q\>\<p,q|[i\hbar(\d/\d t)-\H(P,Q)]|p',q'\> \no \\
     \times\<p',q'|\Psi(t)\>\;dp\,dq\;dp'dq'/(2\pi\hbar)^2 \;,\en
     which defines two action functional integrands, one for the classical action functional, an `island' surrounded  
     by a larger `main-land', representing the quantum action functional, and where each `speck of soil' represents a unique, normalized, Hilbert space vector. On the `island', there is a `flag-pole on which the flag $|p,q\>$'  is displayed, and on the `main-land' there is a `flag-pole on which the
     flag $|\Psi\>$' is displayed. Although the `classical-land' can be reached more easily from the `bridge', the author wanted to show that a very similar procedure may be used to reach either 
     the `classical-land' or the `quantum-land'.\footnote{This story has covered `canonical 
     classical-land', but there are two other `bridges' that reach two other `islands', one for 
     `spin classical-land' and the other for `affine classical-land', both of which are detailed 
     in following sections. It is rumored that the `spin flag is $|\theta,\vp\>$' on their `island', and the `affine flag is $|p;q\>$' on their `island'. Unfortunately, these other `islands' are well off to each side and not displayed in Figure 1.}
     

\begin{figure}[hbt]
  \includegraphics[width=0.95\textwidth]{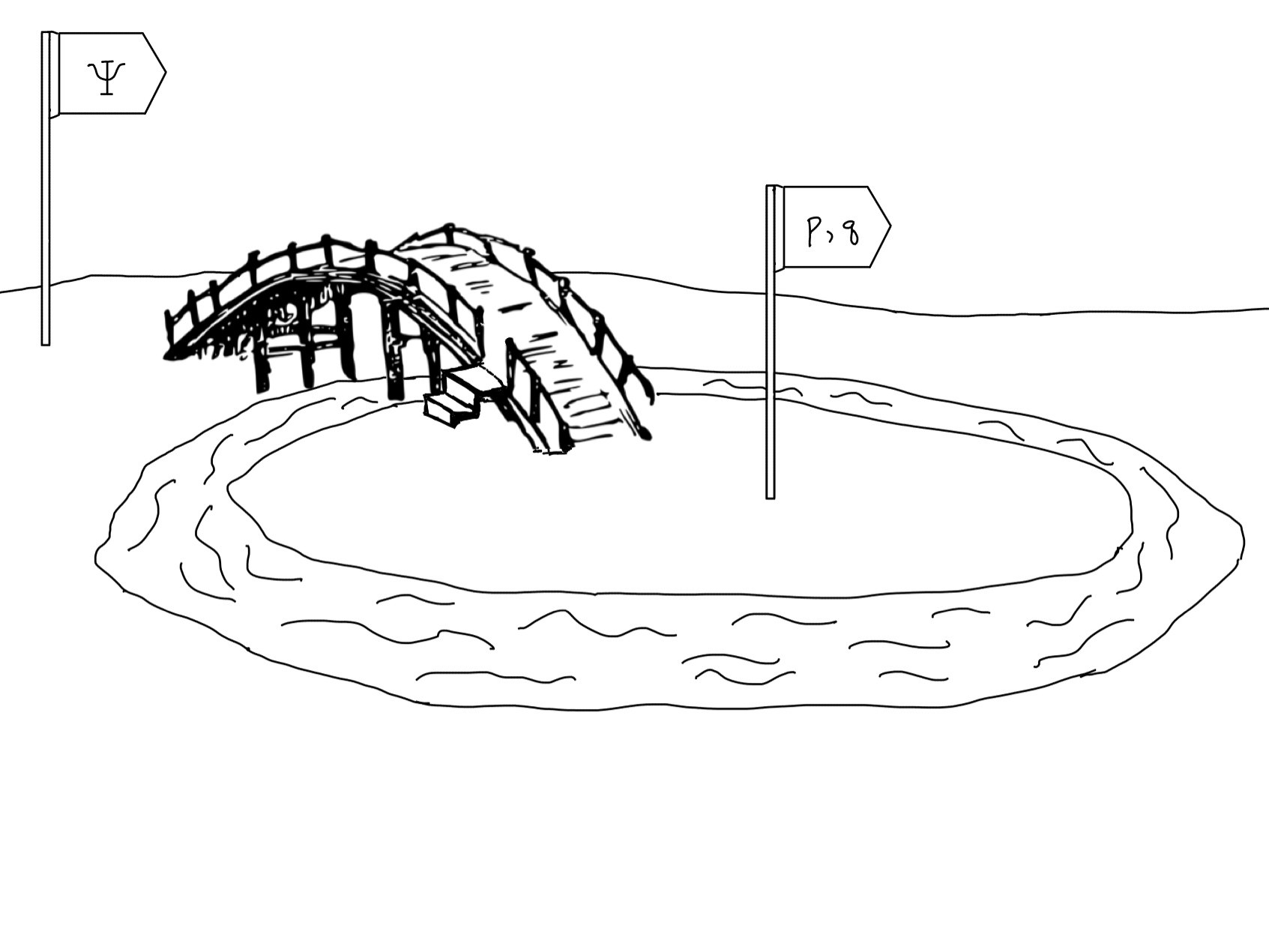}
  \caption{A representation of the `bridge' connecting `quantum-land' with `classical-land'.}
\end{figure}

\section{Classical and Quantum Stories for \\Additional Quantization Processes}
In the analysis in several previous sections we relied heavily on coherent states. To analyze 
the major tasks in this section it is helpful to initially find suitable coherent states. These coherent states rely on seeking suitable basic quantum operators. 

\subsubsection{Spin coherent states}
For our spin quantization story we will involve three operators, namely $S_1, S_2$, and $S_3$, where $[S_1, S_2]=i\hbar \,S_3$, as well as cyclic rotations of the operators. The spin coherent states, for $SU(2)$ and $SO(3)$, are chosen as
    \bn  |\theta, \vp\>\equiv e^{-i\vp S_3/\hbar} e^{-i\theta S_2/\hbar}\,|s,s\>\;, \label{93} \en
    where        $-\pi<\vp\leq\pi$ and 
    $-\pi/2\leq\theta\leq\pi/2$, and the normalized fiducial vector,  $|s,s\>$, has the highest eigenvalue for $S_3|s,m\>=m\hbar|s,m\>$, where $m\in \{ -s, ..., s-1,s\}$, 
  and $S_1^2+S_2^2+S_3^2=\hbar^2\,s( s+1)1\!\!1_s$. Here $s\in \half\{1,2,3,4,...\}$,
    and the resolution of the identity \cite{6} is given by 
    \bn 1\!\!1_s=\int |\theta,\vp\>\<\theta,\vp| (2s+1) \sin(\theta) \,d\theta\,d\vp/4\pi\;.\en
    
    \subsubsection{Affine coherent states}
For our affine quantization story we will involve two operators, namely
$Q$ and $D$, where $0<Q<\infty$ and $[Q,D]=i\hbar\,Q$. Since $Q>0$ the operator $P$ cannot be self adjoint and it is replaced by $D=(PQ+QP)/2$,
and both $Q$ and $D$ can be self adjoint. The coherent state parameters are $-\infty<p<\infty$ and $0<q<\infty$, and, for simplicity, we choose $q$ and $Q$ to be dimensionless; hence $p$, $D$, and
$\beta$, below, 
have the dimensions of $\hbar$. The affine coherent states are chosen as
  \bn |p;q\>\equiv e^{ipQ/\hbar} e^{-i\ln(q)D/\hbar}\,|\beta\>\;. \en 
  In this case, the normalized fiducial vector is chosen as $|\beta\>$, where $ [(Q-1)+iD/\beta]|\beta\>=0$, which implies that $\<\beta|Q|\beta\>=1$ and $\<\beta|D|\beta\>=0$. In this case, the resolution of the identity \cite{6} is given by
  \bn 1\!\!1=\int |p;q\>\<p;q|\,\{[1-\hbar/2\beta]/2\pi\}\;dp\,dq/\hbar \;,\en
  provided $\beta>\hbar/2$.
 
 \subsection{Additional quantum  to classical and \\classical to quantum stories}
 \subsubsection{The spin story}
 
The classical spin Hamiltonian is $G(\theta,\vp)$ and the classical spin  action functional is given by 
     \bn A_{c}=\tint_0^T \{ s\hbar\,\cos(\theta(t))\,\dot{\vp}(t)-G(\theta(t),\vp(t))\}\;dt\;. \en
     Using the facts that $\<s,s|S_3|s,s\>=s\hbar$ and $\<s,s|S_1|s,s\>=\<s,s|S_2|s,s\>=0$, the
     affine spin semi-classical action functional is given by
     \bn &&A_{sc}=\tint_0^T \{\<\theta(t),\vp(t)|[i\hbar(\d/\d t)-\G(S_1,S_2,S_3)]|\theta(t),\vp(t)\>\; dt \no \\
     &&\hskip1.77em=\tint_0^T \{ \<s,s|[\cos(\theta(t))\,\dot{\vp}(t)S_3 +
     \dot{\theta}(t)S_2 \no \\
     &&\hskip4em-\G(\cos(\vp(t))\cos(\theta(t))S_1 +\cos(\vp(t))\sin(\theta(t))S_3-\sin(\theta(t))S_2, 
     \no \\ &&\hskip4em \cos(\vp(t))S_2 -\sin(\vp(t))\cos(\theta(t))S_1 +\sin(\vp(t))\sin(\theta(t))S_3, \no\\
     &&\hskip4em \cos(\theta(t))S_3 -\sin(\theta(t))S_1)]|s,s\>\}\;dt 
     \no \\ &&\hskip1.78em=\tint_0^T \{s\hbar\,\cos(\theta(t))\,\dot{\vp}(t)- G(\theta(t),\vp(t))\}\; dt\;.\en
     
     The passage from semi-classical spin functional and twice applying the two pieces of the identity, and then reuniting the two separate parts together, as was done for the canonical story,
     leads to the quantum affine functional.
     
 \subsubsection{The affine story}
 The classical affine Hamiltonian is $H'(pq,q)$, where $0<q<\infty$ and $-\infty<pq<\infty$, and
 the classical affine action functional is given by 
   \bn A_c =\tint _0^T\{ -\dot{p}(t) q(t)- H'(p(t)q(t), q(t))\}\; dt \;. \en
 The quantum affine Hamiltonian is given 
 by $\H'(D,Q)$, where $Q>0$ and the dilation operator $D=(PQ+QP)/2$. The
 quantum action functional is given by\
     \bn A_{q}=\tint_0^T\<\Phi(t)|[i\hbar(\d/\d t)-\H'(D,Q)]|\Phi(t)\>\;dt\;.\en
 Finally, the semi-classical affine action functional is given by 
    \bn &&A_{sc}=\tint_0^T\<p(t);q(t)| [i\hbar(\d/\d t)-\H'(D,Q)]|p(t);q(t)\>\;dt \no \\
     &&\hskip1.8em=\tint_0^T\<\beta|[-\dot{p}(t)\, q(t)Q+\dot{q}(t)D/q(t)-\H'(D+p(t)q(t)Q,q(t)Q)]|\beta\>\; dt
     \no \\ &&\hskip1.8em=\tint_0^T \{-\dot{p}(t)q(t)-H'(p(t)q(t), q(t))\}\;dt\;, \en
     thanks to the facts that $\<\beta|Q|\beta\>=1$ and $\<\beta|D|\beta\>=0$.
     
     The passage from semi-classical affine functional and twice applying the two pieces of the identity, and then reuniting the two separate parts together, as was done for the canonical story,
     leads to the quantum affine functional.
     
     L. Gouba has examined an affine quantization of a free particle and an half-harmonic oscillator,
     where both problems require that $0<q<\infty$, and was able to extract the eigenfunctions and eigenvalues for the half-oscillator problem \cite{7}.
     While the eigenfunctions are quite different, the eigenvalues for this example are equivalent to those of a full-harmonic oscillator in that they are equally spaced. This puts these results within the larger set of harmonic oscillator type problems.
     
\subsection{Additional examples of a smooth and combined\\ quantum and classical story}
The article \cite{8} summarizes the author's approach to non-renormalizable covariant scalar fields as
well as his approach to quantum gravity. While a canonical quantization approach to these problems has faced difficulties, an affine quantization has come through with flying colors. Readers of the present paper may have noticed that an affine approach is very similar to a canonical approach to quantization, the only difference between the two procedures is that (favored) canonical variables,
such as $p$ and $q$, are promoted to quantum operators, while (favored) affine variables, such as
$pq$ and $q$, are promoted to quantum operators. In \cite{9}, the reader would find a natural analysis that this choice of basic affine variables to promote to quantum operators makes good sense.

Affine procedures that are applied to the difficult problems being considered here, already have equations, such as their Schr\"odinger equations, which are ready for approximate $-$ or perhaps even exact $-$ solutions to be found. Hopefully, smooth procedures between classical and quantum systems may play a helpful role in further analysis.

\section*{Acknowledgements} The author thanks Jennifer Klauder and Dr. Dustin Wheeler for creating Figure 1.

\end{document}